\documentstyle[aps,prd,epsfig]{revtex}
\makeatletter
\@addtoreset{equation}{section}
\makeatother

\newlength{\dinwidth}
\newlength{\dinmargin}
\begin{document}
\def\beq{\begin{equation}}
\def\eeq{\end{equation}}
\def\beqar{\begin{eqnarray}}
\def\eeqar{\end{eqnarray}}
\def\barr#1{\begin{array}{#1}}
\def\earr{\end{array}}
\def\bfi{\begin{figure}}
\def\efi{\end{figure}}
\def\btab{\begin{table}}
\def\etab{\end{table}}
\def\bce{\begin{center}}
\def\ece{\end{center}}
\def\nn{\nonumber}
\def\disp{\displaystyle}
\def\text{\textstyle}
\def\fs{\footnotesize}
\def\arraystretch{1.2}

\thispagestyle{empty}
\def\thefootnote{\fnsymbol{footnote}}
\setcounter{footnote}{1}

\vspace*{2cm}
\begin{center}
{\Large\bf Photon  and Dilepton Production   }   \\
\vskip 0.2 true cm
{\Large\bf from                 }
\vskip 0.2 true cm
{\Large\bf Hot Out-Off-Equilibrium  Media }
\footnote{ Contribution to the XXXVII-th Cracow
 School of Theoretical Physics, {\it Dynamics of Strong Interactions},
 to appear in Acta.Phys.Pol.}
\end{center}

\vspace*{0.3cm}
\begin{center}
{\bf R.~Baier}$^1$, {\bf M.~Dirks}$^1$  {\bf and K.~Redlich}{$^{1,}$}{$^2$}
\\[.3cm]
$^1${\it Fakult\"at f\"ur Physik, Universit\"at Bielefeld,
D-33501 Bielefeld, Germany}\\
$^2${\it Institute for Theoretical Physics, University of Wroclaw, \\
PL-50204 Wroclaw, Poland}\\
\end{center}

\vspace*{3cm}
\section*{Abstract}
The electromagnetic emissivity from  QCD media  away from equilibrium
is studied in the framework  of closed  time path thermal field theory.
For the dilepton rate a nonequilibrium mesonic medium is considered
applying      finite temperature perturbation theory for
 $\pi -\rho$ interactions.
 The dilepton rate
 is  derived  up to the order $   g_\rho^2$.
 For hard photon production a quark gluon plasma is
 assumed and calculations are
 performed in leading order in the strong coupling constant.
These examples are chosen in order to investigate
the role of possible pinch terms in boson and in fermion
 propagators, respectively.
 The implications of the results for phenomenology are also discussed.

\vfill

\def\thefootnote{\arabic{footnote}}
\setcounter{footnote}{0}
\clearpage

\section{Introduction}

Dilepton  and photon production
is  one of the interesting tools
to study  collective effects in strongly interacting matter produced in
ultrarelativistic heavy ion collisions \cite{shur,mclerran,ruuskanen,redl}.
This is particularly
evident from the recent CERN experimental data   at   SPS
  collision energy measured by the collaborations
CERES/NA45 \cite{ceres}   and HELIOS/3 \cite{helios},
which show  qualitatively different behaviour  of the dilepton
yields in A-A as compared to p-p collisions.
Extensive theoretical studies done over the last years
\cite{dinesh,ko,gery,koch,our1,our2} suggest  that this new
behaviour of the measured dilepton yield could be related with
the thermalization of the medium created in heavy ion collisions.
Also the properties of the measured particle spectra at AGS and
SPS energy seem  to be well explained by the production from the  thermal
source.~\cite{model}

Thermalization of a hadronic medium
should be     a fast process. Recent calculations in
 kinetic theory show
that
only a few elastic particle collisions are  already sufficient
   to maintain
thermal particle spectra.~\cite{Gavin} The  chemical
equilibration, in contrast,
could be much slower as it requires a detailed balance
between different reactions  with particle number changing processes.
Also recent predictions of QCD inspired models like HIJIING \cite{hi}
or Self-Screened Parton Cascade~\cite{ss} lead  to     similar conclusions
for  the partonic medium. These models indicate   that the partonic
medium created initially in ultrarelativistic heavy ion collisions
appeared after very short time in  local thermal equilibrium,
 however, chemical saturation is most likely never achieved before the
system hadronises.

The above discussions thus show that when analysing the properties
of strongly interacting matter as produced in heavy ion collisions
we have  necessarily to take into account  nonequilibrium effects.

In this work we show how nonequilibrium effects could possibly
influence the electromagnetic emissivity from QCD media.
In particular we  discuss soft dilepton production from
$\pi -\rho$ interactions
in an off-equilibrium mesonic medium and   hard thermal photon
production from an  undersaturated quark gluon plasma.

We  apply the real time formalism of field theory with particular
emphasis  on the role and structure of the pinch singularities~\cite{Altherr1,Altherr2}
which appear  in  propagators of particles
in a nonequilibrium  medium.

The paper is organized as follows: in Section.2 we introduce the
Keldysh representation and discuss the structure of  particle  propagators
in nonequilibrium media. In Section.3 the dilepton
rate from pi-rho scattering is derived and discussed. In Section.4
we show that there is  self-consistent dynamical screening
which leads to a finite photon rate in an off-equilibrium quark gluon
plasma.

\bigskip
\bigskip
\section{ {Real Time Formalism and  Particle  Propagators}}
In order to see how  nonequilibrium effects
 modify the structure of the particle propagators
we first summarize  some the important relations of the
real time formulation of finite temperature field theory.

In the context of the closed time path or the Keldysh representation
the bare propagator in equilibrium at finite $T$ acquires a   2$\times$2 matrix structure
of the following general form:\cite{Lands,chou,Eijck}

\beq
D^0(k)=
\left(
\matrix{
\Delta_R
 & 0                      \cr
0       & -\Delta_A                 \cr
}
\right)
\mp 2\pi i \delta (k^2-m^2) \epsilon(k_0)
\left(
\matrix{
n(k_0)
 &  n(k_0)     \cr
 \pm 1+ n(k_0) & n(k_0)     \cr
}
\right) ,
\label{21}
\eeq
where $k=(k_0,\vec k)$,
 $k^2 = k^2_0 -{\vec k}^2$ ;
 $\theta$ denotes the step
function, $m$ is the $T=0$ mass and $n(k_0)$ is the thermal
 distribution function
\beq
n(k_0)={1\over {e^{k_0/T} \mp 1}} ,
\label{22}
\eeq
where the upper (lower) sign in the denominator of the above equation
(similarly as in (\ref{21})) corresponds to bosonic (fermionic ) fields.
The $\Delta_R (\Delta_A)$ are the retarded and advanced vacuum
propagators which for  bosons and fermions are given by

\beq
\Delta_{R (A)}^B (k) = \frac{1}{k^2 - m^2 {+\atop {(-)}} i \epsilon k_0}, ~~~~~
\Delta_{R (A)}^F (k) = \frac{{\not k }  +m
}{k^2 - m^2 {+\atop {(-)}} i \epsilon k_0}.
\label{23}
\eeq

In the expressions for $\Delta_{R(A)} (k)$ the usual limit
$\epsilon \rightarrow 0$ is considered, whenever this limit is defined,
e.g.
\beq
\Delta_R^B (k) - \Delta_A^B (k) = - 2 \pi i \epsilon (k_0)
 \delta (k^2 - m^2) ,
\label{24}
\eeq
 where $\epsilon
(k_0) = \theta (k_0) - \theta (-k_0)$ is the sign function.

To account for
 nonequilibrium effects we follow the approximations described
in detail in \cite{chou}, and more recently in \cite{LeBellac2}.
It amounts only
 to  replace the thermal $n(k_0)$, (\ref{22}), by
their nonequilibrium counter parts, i.e. in general by Wigner
distributions $n (k_0, X)$.
This corresponds in terms of the cumulant expansion to approximate
nonequilibrium correlations by the second cumulant only. As a further
approximation we suppress the possible dependence on the center-of-mass
coordinate $X$, essentially assuming a homogenous and isotropic medium.

Practically as   nonequilibrium distributions for bosons and fermions we take,
\begin{equation}
 n(k_0,X) = \left\{
 \begin{array}{c}
            n_F(|k_0|) \\
  1 -            n_F(|k_0|))
 \end{array} \right.
 \quad\quad
 n(k_0,X) = \left\{
 \begin{array}{c}
            n_B(|k_0|) \\
  -(1+           n_B(|k_0|))
 \end{array} \right.
 \quad\mbox{for}\quad
 \begin{array}{c}
  k_0 > 0 \\
  k_0 < 0 
 \end{array}
 \label{25}
\end{equation}
with the J\"uttner parameterizations
\beq
 n_{B(F)} (|k_0|) \equiv \frac{1}{e^{(|k_0| -\mu)/T} \mp 1} \hat= \frac{\lambda}
{e^{|k_0|/T} \mp\lambda} ,
\label{26}
\eeq
or for the actual estimates just the modified equilibrium distributions
\beq
 n_{B(F)} (|k_0|) \equiv
{\lambda\over {e^{|k_0| /T} \mp 1}} ,
\label{26a}
\eeq
expressed   by introducing the fugacity parameter \cite{groot}
$\lambda \equiv e^{\mu/T}$, which is
assumed to be energy independent. Obviously $\lambda \not= 1$ in case of
(chemical)  nonequilibrium.

We have thus seen that if the separation of the macroscopic scale X
from a fast microscopic one in the sense of the Wigner transform
is valid, then the relevant
 nonequilibrium bare propagator has  formally the same structure
 as the corresponding one in  equilibrium.
 This, however, is not in general true for dressed propagators where
 the appearance of the additional
 out-of-equilibrium pinch singular
term is to be expected
\cite{Altherr1,Altherr2,Bedaque}.

To show     the origin and the possible appearance
 of pinch terms, let us consider for illustration,
the selfenergy one-loop  corrections to the (12) component
 of the scalar propagator.
 Denoting the selfenergy
insertion by the matrix $\Pi_{ab} (p)$ we obtain the improved
$(12)$-propagator using the Dyson equation,
\beqar
i D^*_{12} (p) \sim i D_{12} (p) + \sum_{a,b =1,2}
 i D_{1a} (p) (-i
\Pi_{ab}(p)) i D_{b2} (p) \nn \\
& = & i D_{12} (p) + i \delta D_{12} (p).
\label{28}
\eeqar
With the propagators specified in (\ref{21},\ref{23}) and with the relations
\beq
\Pi_{11} (p) = - \Pi^\star_{22} (p), ~~ {\rm Im}\, \Pi_{11} = \frac{i}{2}
(\Pi_{12} + \Pi_{21}),
\label{29}
\eeq
also valid out-of-equilibrium, we obtain
\beqar
\delta D_{12} (p) &=& n (p_0,X) (\Delta^2_R (p) - \Delta^2_A (p))
{\rm Re} \, \Pi_{11} (p) \nn\\
&+& \frac{1}{2} n (p_0,X) (\Delta^2_R (p) + \Delta^2_A (p))
[ \Pi_{12}(p) - \Pi_{21}(p)] \nn \\
&+& \Delta_R (p) \Delta_A (p)
 [ n(p_0,X) \Pi_{21}(p) - (1 + n (p_0,X)) \Pi_{12}(p) ].
\label{210}
\eeqar
Here the ill-defined product $\Delta_R (p) \Delta_A (p)$ appears, giving
rise to possible unpleasant pinch singularities \cite{Altherr1,Lands}.
In case of equilibrium, however, it is well known that the last term in
(\ref{210})
vanishes due to the detailed balance condition,
\beq
\Pi_{21} (p) = e^{p_0/T} \Pi_{12} (p).
\label{211}
\eeq
As a consequence
 (\ref{210})
 simplifies to
\beq
\delta D_{12}^{eq} (p) =  2\pi i \epsilon (p_0) n (p_0)\left[ \delta^\prime
(p^2) {\rm Re}\, \Pi (p)
+ \frac{1}{\pi} {\bf P} \left(\frac{1}{p^2}\right)^2 {\rm Im}\,\Pi (p) \right],
\label{212}
\eeq
where
\beq
{\rm Re}\, \Pi (p) \equiv {\rm Re}\, \Pi_{11} (p), \, ~~\,
 {\rm Im}\,\Pi \equiv
\frac{1}{2} \epsilon (p_0) \frac{i\Pi_{12} (p)}{n(p_0)} ,
\label{213}
\eeq
 expressed in terms of the equilibrium distribution $n(p_0)$,
and in terms of the (retarded) selfenergy
$\Pi(p_0 + i\epsilon p_0, \vec p)$ \cite{LeBellac}.
$\delta^\prime$ denotes the derivative of the $\delta$-function, and ${\bf P}$
the principal value.

For nonequilibrium distributions (\ref{25}), for which obviously
$(1+n)/n \not= \exp (p_0/T)$, the last term in (\ref{210})
 does not cancel and may give rise to  pinch  singularities,
 which, however, have to be regularized \cite{Altherr2} by proper
 resummation.

The appearance of the  pinch term in the particle propagator
 implies a
 non-trivial modification of different physical
processes which are relevant  in out-of-equilibrium systems.
In the following we  discuss how the nonequilibrium
effects  modify   dilepton and photon rates
in strongly interacting matter. In particular we show
that the pinch (singular) term can be successfully regularized
leading to finite rates for both dilepton and photon production
from a  hot mesonic and quark-gluon medium, respectively.
\bigskip
\section{Dileptons from a nonequilibrium mesonic medium}

As the first example of the application of the real time formalism
to describe   nonequilibrium effects we consider in this section
dilepton production from a  mesonic medium.

\begin{figure}
\centering
\epsfig{file=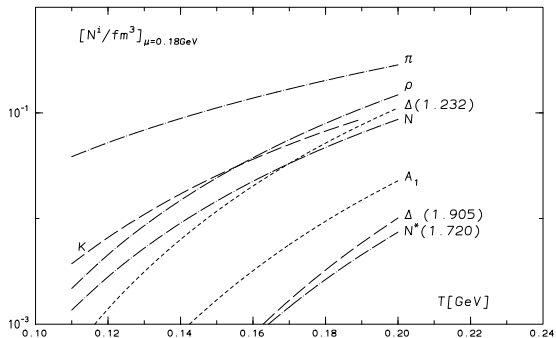,angle=90, width=10cm}
\caption{\label{fig:kr} Particle multiplicity  per fm$^3$
 for a baryon chemical potential $\mu_B=0.18$GeV as a function
 of temperature $T$ }
\end{figure}

The absence of chemical equilibrium in a hadronic  medium can be effectively
taken into account  modifying the particle   distribution function~\cite{Gavin}
by the fugacity parameter  $\lambda$ as explained in the last section.
In addition we       simplify the distribution
adopting  the Boltzmann approximation
$n(k_0,X) \simeq \lambda e^{-|k_0|/T}$, and
assume  that $\delta \lambda \equiv \lambda - 1$
is small $\delta \lambda \ll 1$ , i.e. a situation not far
out-off-equilibrium.

Thermal dileptons are mostly produced from hadronic decays,
particle bremsstrahlung and hadron-hadron scattering.
In Fig.~1 we show different particle number densities as a function
of temperature  for a fixed value of the baryonic chemical potential $\mu_B =0.18$GeV.
As one can see in Fig.~1  pions and rho mesons are the most abundantly
present particles in a medium in the temperature range $0.14<T<0.20$GeV
relevant for heavy ion collisions at SPS energy.
 Thus, discussing soft dilepton production in a nonequilibrium
 hadronic medium we consider first dileptons originating from
 $\pi -\rho$ scattering.

    The interaction of the charged pions with the neutral
 massive rho meson field $\rho_\mu$ and
the  electromagnetic potential $A_\mu$ is described by the
  Lagrangian \cite{kapusta}
\beq
L
= |D_\mu\Phi |^2
 - {1\over 4} \rho_{\mu\nu } \rho^{\mu\nu } +
{1\over 2} m_\rho^2 \rho_{\nu }\rho^{\nu }
-
{1\over 4} F_{\mu\nu } F^{\mu\nu },
\label{31}
\eeq
where $D_\mu\equiv \partial_\mu -ieA_\mu -ig_\rho\rho_\mu$
is the covariant derivative, $\Phi$ is the complex
 pion field in the following assumed to be massless,
$\rho_{\mu\nu }$ is the rho and $F_{\mu\nu}$ is the
photon field strength tensor.
As is well known, the $\rho \pi \pi$ coupling is rather large
$g^2_\rho / 4 \pi \simeq 2.9$; nevertheless we attempt an
effective perturbative treatment.

\begin{figure}
\centering
\begin{minipage}[c]{10cm}
\centering
\epsfig{file=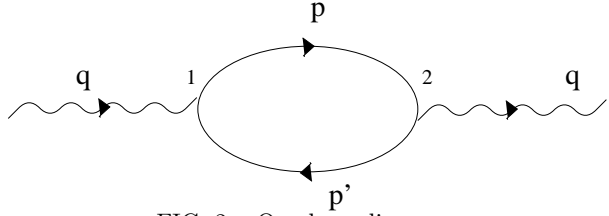, angle=-90, width=8 cm}
\caption{\label{fig:kz1} One-loop diagram
  }
\end{minipage}%
\end{figure}

The thermal emission rate of heavy photons with invariant mass
$M$, energy $E$ and momentum $\vec q$ can be obtained from  the trace of
the photon selfenergy tensor $\Pi_{\mu\nu}^{\gamma}$ as follows~\cite{mclerran,weldon}
\begin{equation}
{{dR}\over {dM^2d^3q/E}} =
{\alpha\over {48\pi^4 M^2}}{}i\Pi_{12}^{\gamma} (q_0, \vec q ).
\label{32}
\end{equation}

The virtual photon selfenergy is usually approximated
by carrying out a loop expansion to some finite order.
 From the Lagrangian (\ref{31}) and using the
   closed-time-path formalism \cite{Lands,chou,Eijck}
 we calculate  the dilepton rate          in a
 thermal pionic medium at the two-loop level.
Typical diagrams are shown in Figs.~2,3 and 4.
The two tadpole diagrams of $O(g_\rho^2)$ are not included,
since they do not contribute to the discontinuity of the
photon selfenergy (\ref{32}).


%
\begin{figure}[ht]
\centering
\epsfig{file=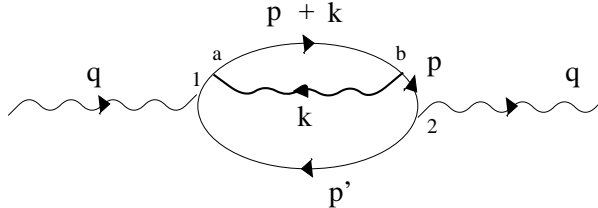, angle=-90, width=8cm}
\caption{\label{figub}\it{Two-loop diagram: selfenergy insertion.
The labels a,b =1,2 denote the type of $\pi \rho$ vertex.
The solid line with momentum label k corresponds to the $\rho$}}
\end{figure}
On the one loop level,  Fig.~2, the photon selfenergy
is expressed  by the tree level particle propagators.
Thus, the only difference with equilibrium calculations enters
through the distribution function. In  a rather straightforward way
one recovers the standard expression for the Born term,
$\pi^+\pi^-\to \gamma^* \to e^+e^-$,~\cite{mclerran}, which
 under the Boltzmann approximation and
 for  $M<m_\rho$ reads:
\begin{equation}
{{dR}\over {dM^2{{d^3q}\over {E}}}}\simeq (1+2\delta\lambda)
{{\alpha^2}\over {96\pi^4}} |F_\pi(M)|^2 \exp (-E/T)
\label{35}
\end{equation}
where $|F_\pi|$ is the electromagnetic pion form factor and
$E^2={\vec q}^2+M^2$.

The only difference of the above result with the one
derived in the equilibrium medium is the appearance of
the $(1+2\delta\lambda)$ factor which accounts for nonequilibrium effects.
The more complete expression for the Born contribution
with quantum statistics (c.f. (\ref{25})) and for arbitrary heavy photon momentum
is, however, also  known     in the literature (see \cite{redl}).

For soft dileptons $M<m_\rho$ we consider in the following the mass
distribution per unit space-time volume $dN/dM^2d^4x$.\cite{our1}
To go beyond the one-loop approximation and to include the contributions
due to $\pi^\pm -\rho^0$ scattering one needs to
calculate the two loop diagrams shown in  Figs.~3-4.
 There are two types of            diagrams
contributing to the thermal dilepton rate from $\pi -\rho$ interactions. These are the diagrams
with real and virtual $\rho^0$ vector mesons.
The processes involving real  ${\rho^0}^,$s  are due to
$\pi\pi\to\rho\gamma^*$, $\pi\rho\to\pi\gamma^*$ and $\rho\to\pi\pi\gamma^*$
reactions.
 These contributions are obtained by cutting the two-loop
diagrams (Figs.~3 and 4) such that the $\rho$ is put on-mass shall.
  The virtual rho contributions, obtained by cutting the diagrams shown
  in Figs.3-4
  lead to $O(g_\rho^2)$ order corrections to the Born term.
\begin{figure}[ht]
\centering
\begin{minipage}[c]{7cm}
\centering
\epsfig{file=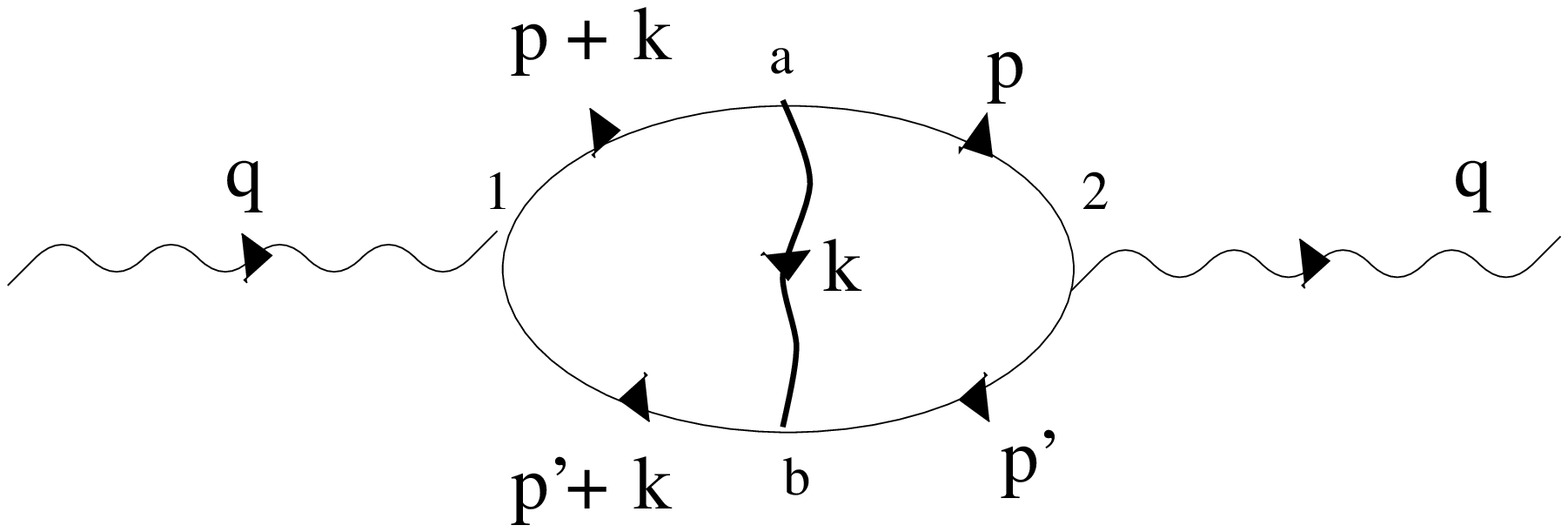, angle=-90, width=7cm}
\end{minipage}%
\vskip .7cm
\begin{minipage}[b]{6.0cm}
\centering
\epsfig{file=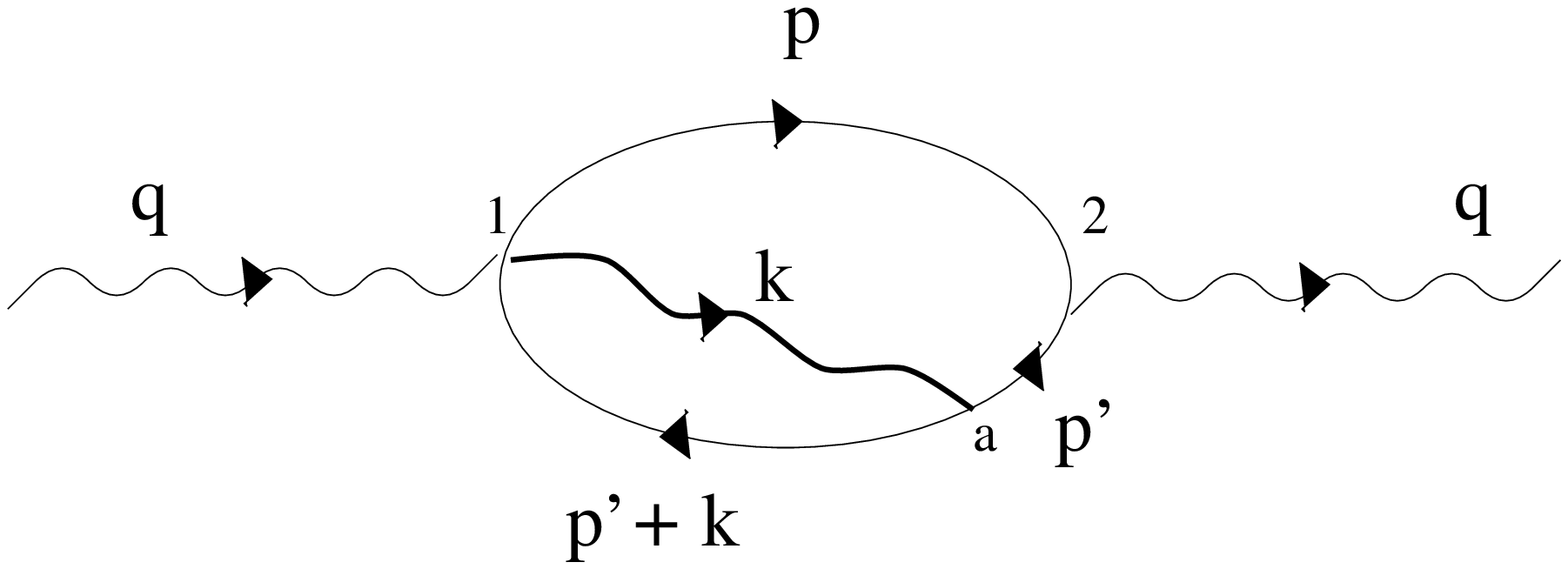, angle=-90, width=6.0cm}
\end{minipage}%
\hspace{1cm}
\begin{minipage}[b]{6cm}
\centering
\epsfig{file=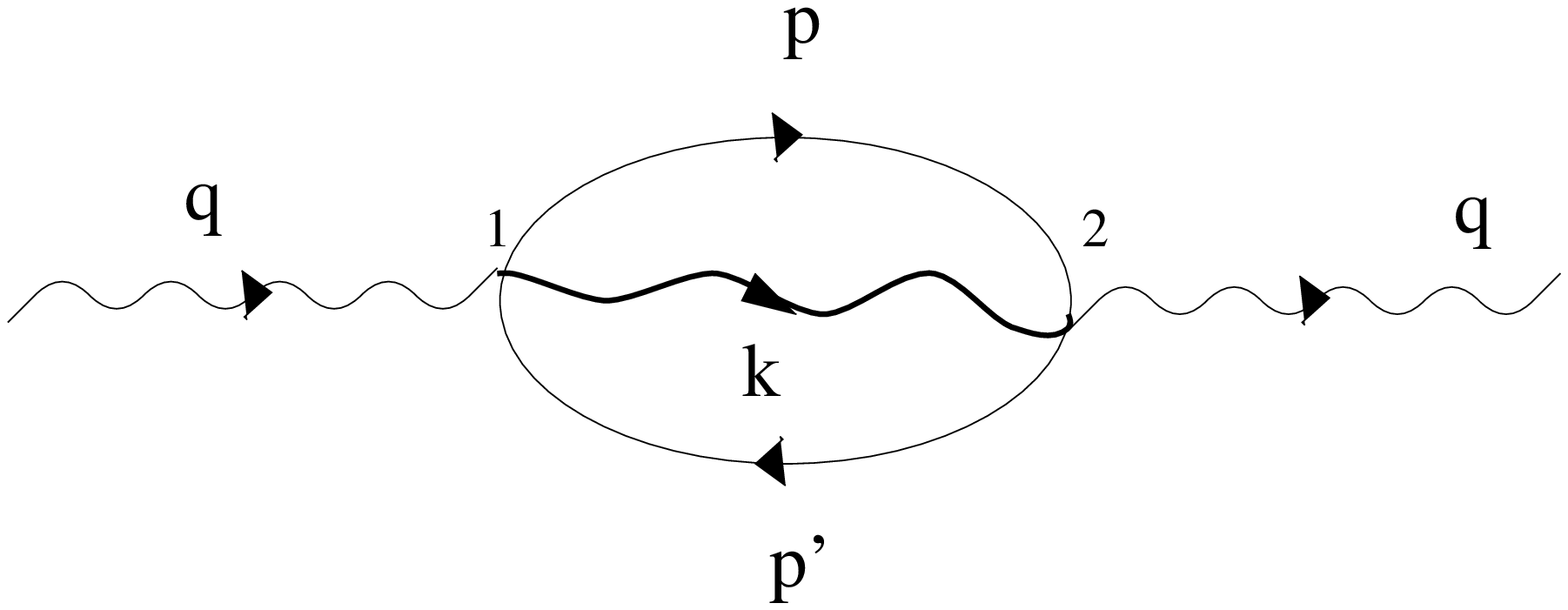, angle=-90, width=6cm}
\end{minipage}
\vskip .7cm
\caption{\label{fig4}\it{Two-loop diagrams: vertex type corrections}}
\end{figure}
  In the following we will discuss real and virtual rho contribution separately.
\bigskip

\subsection{Dilepton rate with on-shell rho contribution}

We start to discuss explicitly the contribution from the diagram shown
in Fig.~3, where the selfenergy correction $\delta D_{12} (p)$ is inserted into
the pionline with momentum $p$. The expression for $\delta D_{12} (p)$ is
derived in Section 2  using the Dyson equation.

Due to the structure of the (one-loop) selfenergy correction to the pion
propagator (\ref{210}) a more careful derivation is required.
 Using  the second and the third term of (\ref{210}) we calculate the "real" correction
to $\Pi_{12}^\gamma$,
\begin{eqnarray}
 i\delta \Pi_{12}^{\gamma,off}(q^0,\vec q)& &
 \simeq - (-ie)^2 \int\frac{d^4 p }
 {(2\pi)^4}\; (p+p')^2 \left\{ {\bf P}(\frac{1}{p^2})^2 \; n(p_0,X) \right.
 \, (i \Pi_{12}(p) - i\Pi_{21}(p) ) \nonumber\\
 & &+ \left. \frac{1}{(p^2)^2 + (p_0\gamma)^2}
 \,(n(p_0,X) i \Pi_{21}(p) - (1+n(p_0,X))i \Pi_{12}(p) ) \right\}
 i D_{21}(p^,),
\label{41}
\end{eqnarray}
where
a corresponding term with $(p \leftrightarrow p')$ has to be added.
The pion selfenergy $i \Pi_{12} (p)$ at one-loop order due to
$\pi\rho$ interactions (\ref{31}) is given by
\beq
i\Pi_{12} (p) =
 g^2_\rho \int \frac{d^4k}{(2\pi)^4} \, (2p+k)^\sigma \left(-g_
{\sigma\tau} + \frac{k_\sigma k_\tau}{m^2_\rho} \right) (2 p + k)^\tau
\, D_{12} (p+k) D_{21} (k).
\label{42}
\eeq

In the second term in (\ref{41}) one can recognize  the contribution of the
 pinch term proportional to the product $\Delta_R\Delta_A$,
 which is not cancelled in case of nonequilibrium
distributions \cite{Altherr1}.
To
  regularize  \cite{Altherr2} the pinch singularity we have explicitly included
  a finite damping
rate $\gamma$
of the pion.

The pion damping rate $\gamma$ is determined by the ``pole'' (in the lower
energy half-plane) of the retarded propagator. In the following
  we        neglect  thermal corrections due to the
real part of the (retarded) pion selfenergy, such that
\beq
\gamma \simeq - \frac{{\rm Im}\, \Pi (p_0,
{\vec p})}{p_0} ,
\label{43}
\eeq
evaluated on-shell $p^2 = 0$ for positive pion energy $p_0$. In the
one-loop approximation the dominant contribution
to the absorptive part of the
pion selfenergy
 comes from the pions
thermal distribution. In the Boltzmann approximation it leads to
\beq
{\rm Im}\, \Pi (p_0 = p , p) \simeq - \frac{g^2_\rho}{4\pi}
\frac{m^2_\rho}{4p} \int^\infty_{\frac{m^2_\rho}{4p}} n (E_\pi,X) d E_\pi
\simeq - \lambda
\frac{g^2_\rho}{4\pi} \frac{m^2_\rho T}{4p} e^{-\frac{m^2_\rho}{4pT}} .
\label{44}
\eeq
${\rm Im}\, \Pi$ vanishes for ${\vec p} = 0$, and it has its
maximum near $p\simeq \frac{m^2_\rho}{4T}$. We note that (\ref{44}) gives
a positive damping rate $\gamma$, as required. In order not to
overestimate the contributions arising from the $\Delta_R\Delta_A$ terms we
take for the numerical estimates a momentum independent value,
namely the one at the maximum,
\beq
p_0 \gamma \simeq - {\rm Im}\, \Pi\,(p_0 \simeq p, p) \simeq
\lambda \frac{g^2_\rho}{4\pi}\frac{1}{e} T^2 .
\label{45}
\eeq

We discuss further  the estimate for the dominant contribution
which is due to $\pi \rho \to \pi \gamma^\star$ in the limit $\gamma \to
0$.
In the Boltzmann approximation
the  contribution arising from (\ref{41}) to the
dilepton spectrum reads \cite{our1}
\begin{eqnarray}
& &\; \; \frac{dN^{pinch}}{dM^2 d^4 x } \stackrel{\gamma \to 0}
   {\simeq} -~ \delta \lambda
 \,\frac{\alpha^2 (g_\rho^2/4\pi) }{48 \pi^4 M^2} \sqrt{\frac{\pi T^3}
 {2 m_\rho^3}}  \nonumber\\
 &  \times & \left\{ \int_{m_\rho^2}^{2m_\rho^2 + M^2}   du \;
  e^{- \frac{u+m_\rho^2}{ 2m_\rho T}}
  \int_{t_{min}}^{t_{max}}dt + \int_{2m_\rho^2 + M^2}^\infty du \;
 e^{-\frac{u+m_\rho^2}{2m_\rho T}}\int_{-m_\rho^2}^{t_{max}} dt \right\}
 \;
 \frac{m_\rho^2M^2 }{t^2 + \gamma^2 }~,
  \label{46}
\end{eqnarray}
 When using in (\ref{46}) the momentum
independent
value for the damping $\gamma$  (\ref{45}), regularizing pinch
singularities at  $t^2= 0$,
the integrations can be explicitly
performed. The leading term with $\gamma \to 0$ is proportional to $1/
\gamma$, indicating the pinch singularity,
\begin{equation}
 \frac{dN^{pinch}}{dM^2 d^4 x}\simeq
- ~{{\delta \lambda}\over \lambda} \;
 \frac{\alpha^2 (g_\rho^2/4\pi)}
 {24 \pi^3} \sqrt{\frac{\pi T^3}{2m_{\rho}^3}}
\; \frac{m_\rho^3 }{g_\rho^2/
 (4\pi e) T} \; e^{-\frac{2m_\rho^2 +M^2}{2m_\rho T}} ,
 \label{47}
\end{equation}
which is actually independent of the coupling $g_\rho^2$.

To obtain   the overall rate for dilepton production with real (neutral)  rho
contributions one still needs to calculate the first term in
(\ref{41}). According to the  calculations
presented in \cite{our1}, in the limit of soft dilepton mass $M<m_\rho$, this term reads
\beqar
 & &\frac{dN^{{\rm real}}}{dM^2 d^4 x}  \simeq
\frac{\alpha^2 g^2_\rho / 4\pi}{24 \pi^4 M^2} \sqrt{\frac{\pi T^3}{2 m_\rho^3}}
\nn \\
&  &\times  \left\{ \left[ (1+2\delta\lambda )\int^\infty_{(m_\rho + M)^2} d s \, e^{-
\frac{s+m_\rho^2 - M^2}{2m_\rho T}} + (1+\delta\lambda ) \int^{(m_\rho - M)^2}_0 ds \,
e^{-m_\rho / T} \right] \int^{u_+}_{u_-} du  \right. \,
\nn \\
&  & +  \left. (1+2\delta\lambda )2\, {\bf P} \int^\infty_{m_\rho^2} dt \, e^{-\frac{t + m_\rho^2}
{2m_\rho T}} \int^{u_{{\rm max}}}_{u_{{\rm min}}} du \right\} \nn \\
&  & \times  \left[ 2 + \frac{m_\rho^2 M^2}{4} \left( \frac{1}{t^2} +
\frac{1}{u^2} \right) + \frac{(m_\rho^2 + M^2)^2 + m_\rho^2 M^2/2}{tu}
 -
 (m_\rho^2 + M^2) \left( \frac{1}{t} + \frac{1}{u} \right) \right],
\label{49}
\eeqar
where
\beqar
 & & u_{\buildrel +\over{(-)}} = {1 \over 2} \, (m_\rho^2 + M^2 -s)\,
{\buildrel +\over{(-)}} \, {1 \over 2} \,
  \sqrt{(s-(m_\rho + M)^2) (s-(m_\rho - M)^2)} ,
 \nn \\
& &  u_{{\rm max}} = m_\rho^2 M^2 / t , \, ~~
 u_{{\rm min}} = m^2_\rho + M^2 - t ,
\label{48}
\eeqar
with $ ~s+ t + u = m_\rho^2 + M^2$.
The $u$-integration may still be performed analytically.

In the above formula we have      included  the corresponding
contributions to the real $\rho^0$ processes from the two-loop vertex type
diagrams of Fig.~4.  The first integral in (\ref{48}) corresponds
to $\pi\pi\to\rho\gamma^*$,
the second to
 $\rho\to\pi\pi\gamma^*$
and the third to $\pi\rho\to\pi\gamma^*$ process.

 Finally   the  total nonequilibrium rate
 from real rho processes is obtained as a sum of (\ref{48}) and the pinch term (\ref{47}).
\bigskip

\subsection{Higher order virtual corrections}

For heavy photon production at order $O(g_{\rho}^2)$  we have,
in order to correct the Born rate, to include
virtual $\rho$ contributions, which arise from the processes shown
in Figs.~3 and 4 by cutting the diagrams in the proper way
only through pion lines, without cutting the $\rho$ line.

In some detail we describe the contribution due to the
 selfenergy diagram (Fig.~3).
Here we take the first term in  (\ref{210})
 proportional
to the real part of the pion selfenergy
 $\Pi (p)$ and evaluate

\beq
i \delta \Pi^{\gamma,virtSE}_{12} (q^0 , \vec q )  =
 (-ie)^2 \int \frac{d^4 p^\prime}
{(2\pi)^3} \, (p + p^\prime )^2 \, \epsilon (p_0)~ n(p_0,X)
\, \delta^{\prime}({p^2})  \, {\rm Re} \Pi(p) \, i D_{21} ( p^\prime ).
\label{51}
\eeq

The temperature-independent part of ${\rm Re} \Pi$ is
 absorbed in the definition of
the $T = 0$ pion mass (which we take approximately as $m_{\pi} = 0$).
The temperature dependent part in the one-loop case under
consideration is  weighted by the thermal distribution either for the
$\rho$ meson or for the pion (see \cite{song}). The first case is
of $O(\exp(-m_\rho/T))$, i.e. negligible, the second one therefore
dominates and is expected to be of $O(T^2/m_{\rho}^2)$, due to the presence
of the $T = 0$ $\rho -$ propagator in the loop (Fig.~3).

In the following we present the  result for the virtual rho
contribution  obtained from the selfenergy diagram Fig.~3.
In the limit $m_\rho \gg T$,
having in mind the mass region $M \le m_\rho$ we get for (\ref{51})
\beq
i \delta \Pi^{\gamma,virtSE}_{12} (M , \vec q = 0 )  \simeq
 -\lambda {{e^2} \over {8 \pi^2}}~  \frac{g_\rho^2}{4\pi}
~M^2~ {{T^2} \over {m_\rho^2}} \exp ( - M/T)~.
\label{52}
\eeq

In an analogous treatment     the virtual $T -$ dependent
contributions from the vertex type diagrams, namely from the
first two of Fig.~4 have been      calculated in \cite{our1}.
The final results for pion annihilation including   virtual
rho contributions may be estimated as,
\beq
  \frac{dN^{{\rm Born + virtual}}}{dM^2 d^4 x}
  \simeq
   \frac{dN^{{\rm Born}}}{dM^2 d^4 x}
\left[~ 1 - \lambda\frac{7}{\pi}~ \frac{g_\rho^2}{4 \pi}
 ~(\frac{T}{m_\rho})^2 ~ \right]~
 \label{vir}
\eeq
valid for $m_\rho \gg T$, and for $M < m_\rho$. It is interesting
to note that the $O(g_\rho^2)$ T-dependent corrections (\ref{vir})
are negative and could be large. It suggests to perform resummations.
\bigskip

\subsection{Dileptons from a nonequilibrium  medium - quantitative  discussion}

In the last sections we have shown that the  real time
formalism of finite temperature field theory is an adequate tool to
consistently describe dilepton production from $\pi-\rho$
scattering in an off-equilibrium mesonic medium. In the following we discuss
quantitative  properties of the dilepton rate with the
emphasis on the importance of   nonequilibrium effects.

\begin{figure}
\centering
\epsfig{file=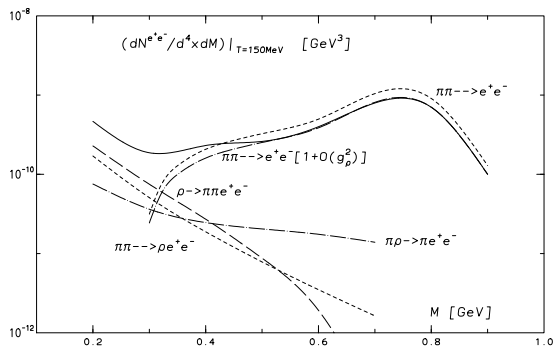,angle=90,width=10cm}
\caption{\label{fig:k12} Dilepton rate as a function of $M$
 at fixed $T$=0.15GeV. The solid line represents the sum of all
contributions }
\end{figure}

In Fig.~5  we calculate       the over all dilepton rate from $\pi-\rho$
scattering in an equilibrium medium that is for $\lambda =1$. We indicate in Fig.~5
 the contributions from different
processes originating from            the  two-loop approximation
of the virtual photon selfenergy.
As expected the $\pi^+\pi^-\to \rho^0 e^+e^-$
reaction is a dominating source of very soft dielectrons with invariant
mass $M<0.3$GeV. Thus, it could be a competing process  with
pion bremsstrahlung and Dalitz decay.
At higher invariant mass $M>0.5$GeV the reactions with the on-shell rho
meson in the initial state $\pi\rho\to \pi\ e^+e^-$ and $\rho\to \pi\pi\gamma^*$ are
 dominating processes calculated
at the two-loop level. However, these contributions are still  much smaller than the
   Born term due to pion  annihilation
 $\pi^+\pi^-\to\rho\to\gamma^*$
   obtained with the effective vertex from VDM.
   This is mostly because of the
 pion form factor which near      the rho resonance increases the rate
 by
 more than an order of magnitude. The destructive interference
 of the amplitudes in the different channels and also smaller
 phase space are the additional   reasons for this behaviour.

 The contribution  of the virtual rho meson calculated
 at    $e^2g_\rho^2$ order should be considered as finite-T corrections
 to the Born
   $\pi^+\pi^-\to\rho\to\gamma^*\to e^+e^-$  annihilation process.
   As seen in Fig.~5 these corrections are  negative and rather large.

In order to analyse the nonequilibrium  effects on dilepton production
from the  mesonic medium
we plot in Fig.~6 the total rate
 characterized by different
values of the  chemical potential $\mu$. We observe that
for positive $\mu$
the rate is increasing with increasing $\mu$. However,
naively one would expect an increase by a factor of $\lambda^2$,
whereas the results in Fig.~6 show a much lower enhancement.
Changing  $\mu $ from $\mu =0$ to $\mu =100$ MeV only an effective
increase by a factor 2 results in Fig.~6, contrary to a factor
4 expected from  $\lambda^2$. This is mostly because of the negative contribution of
the $pinch-singular$ term    summarized in (\ref{47}). For negative values of $\mu$,
that is for $\lambda<1$ the pinch term (\ref{47}) is positive, thus it increases
the contribution of  the $\pi\rho\to\pi\gamma^*$
process in the soft part of the dilepton spectrum. Consequently for $\lambda <1$
 the spectrum is much flatter, with a less
pronounced rho meson peak, as seen in Fig.~6.
From this we deduce the importance of taking
into account the non-trivial pinch term (\ref{47}), which is due to
the structure of the pion propagator in a nonequilibrium medium.

The possible phenomenological implications of the above results,
in  particular the role of  $\pi -\rho$ scattering
in the description of the recent CERN  experimental data on soft dilepton
production,  have been discussed in \cite{our1,our2}
\bigskip

\begin{figure}
\centering
\epsfig{file=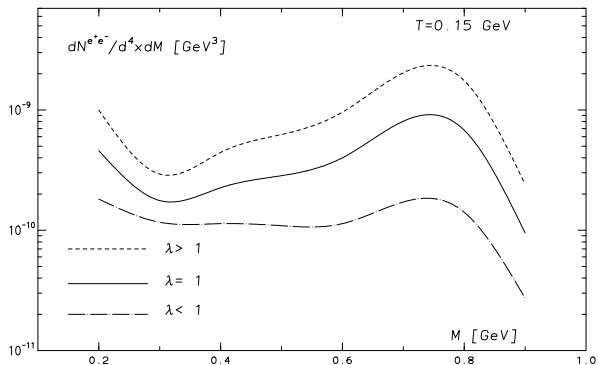,width=8cm}
\caption{\label{fig:k15} Dilepton rate due to $\pi -\rho$
interactions for different values of the nonequilibrium chemical
potential}
\end{figure}

\section{Hard thermal photons  from nonequilibrium quark-gluon plasma}

In this section we present the second example
of applications of the real time formalism to describe
 the production
of hard photons from an  off-equilibrium quark-gluon plasma (QGP).

The photon yield       is one of the possible observables to signal
a deconfined medium produced in  ultrarelativistic
heavy ion collisions. To  lowest order in strong interactions hard
photons in a QGP  are predominantly produced from annihilation and Compton processes,
\begin{equation}
 q + \bar q \to g + \gamma, \qquad
 q (\bar q) + g \to q (\bar q) + \gamma.
 \label{61}
\end{equation}
\begin{figure}
\centering
\epsfig{file=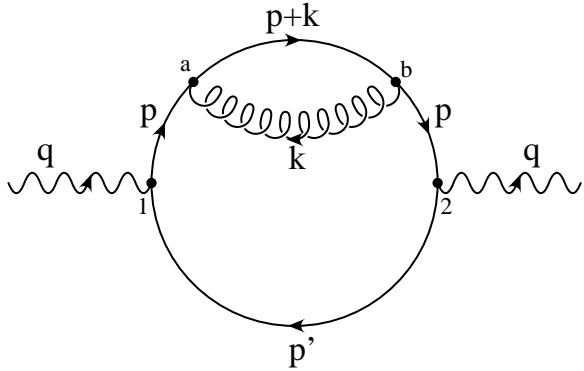,width=8cm}
\caption{\label{fig:fix} The photon polarization tensor $\Pi_{12}$
 to first order in  $\alpha_s$ for
 the real photon production. The momentum labels define the notation used
 in the text. External vertices are to be of type 1 and 2, respectively,
 internal vertices are to be summed over $a,b = 1,2$ }
\end{figure}
It is thus clear that the photon rate is very sensitive to the momentum distribution
of quarks and gluons. Recently different
QCD motivated phenomenological models \cite{hi,ss} suggest that the partonic
medium created initially in heavy ion collisions
reach local thermal equilibrium after very short time,
 however, being
far away from chemical saturation. Thus, similarly as we have done
for the mesonic medium (see (\ref{25}), we  assume that parton distributions
can be approximated by thermal phase space distributions, however,
with nonequilibrium fugacities.

The rate of photon emission can be derived (c.f. \ref{32})
\begin{equation}
 E\frac{dR}{d^3 q} = \frac{i}{2(2\pi)^3} {\Pi_{12}}_\mu^\mu (q),
 \label{62}
\end{equation}
from the trace of the (12)-element $\Pi_{12}$ of the photon-polarization
tensor
calculated  to leading order in the strong
interaction of quarks and gluons. The relevant diagram in fixed order
perturbation theory is shown in Fig.~7. The relevant (12)
and (21) propagator components depend
on the (off-equilibrium) distributions of quarks and gluons.

For real photons the selfenergy $\Pi_{12}$ in (\ref{62}) receives
contributions only from spacelike loop-momenta $p^2  \le 0$.
The fixed order result from the
diagram Fig.~7 turns out to contain IR singular contributions
from small $p^2 \to 0$. For the equilibrium case it has been shown that
this unphysical behavior can be cured taking  hard thermal loop
(HTL) contributions \cite{HTL}
from all
orders of the quark selfenergy consistently into account
\cite{kapusta,rb}.
In the spirit of the HTL-resummation program it is important to distinguish
regions of hard from those of soft momenta \cite{braaten}.
We choose to separate
these scales along the line $p^2 = - k_c^2$ with $k_c^2$ to be chosen on
an intermediate scale:
\begin{equation}
  -p^2_{hard} \sim T^2 \ge  k_c^2 \sim gT^2 \ge -p^2_{soft} \sim g^2 T^2.
\end{equation}
For soft momenta resummation of leading HTL-contributions from
one-loop selfenergy insertions is accomplished by substituting the effective
resummed propagator as indicated by a  blob in Fig.~8.
In an equilibrium medium
the resulting partial rate is IR finite and is shown to exactly match the
hard contribution thereby introducing the thermal mass parameter $m_q$ as an
effective IR cut-off.

Turning to the off-equilibrium situation now our first task is to
determine the appropriate generalization of the effective quark
propagator on the hard and the soft scale.

\subsection{Quark propagator -   hard momentum scale}
\bigskip
For hard momentum $-p^2 \ge k_c^2$ only the one-loop selfenergy correction
to the quark propagation is to be considered. Denoting
the quark selfenergy insertion by $\Sigma$ the  quark propagator
can be obtained from the Dyson equation similar to (\ref{28}),
but in terms of the quark distribution $n_q(p_0,X)$,
\begin{eqnarray}
 \delta S_{12} (p) &=&
 -   n_q(p_0,X) (\Delta_R^2 -  \Delta_A^2) \not p Re \Sigma(p) \not p
 - \frac{1}{2}  n_q(p_0,X) (\Delta_R^2 + \Delta_A^2)
 \not p  (\Sigma_{12} - \Sigma_{21}) \not p \nonumber\\ \label{eq:delta}
 & & - \Delta_R \Delta_A \not p
  [ (1- n_q(p_0,X)) \Sigma_{12} +  n_q(p_0,X) \Sigma_{21} ] \not p.
 \label{63}
\end{eqnarray}
The first term in this expression corresponds to virtual corrections
 and vanishes for real photon production.
The third genuine off-equilibrium contribution, give rise to pinch singularity.
However,  due to the restricted kinematics $-p^2 \ge k_c^2$ the singularity
at $p^2 =0$ is never crossed so that in this
region the pinch combination is equivalent to the principal
value appearing in the second term:
\beq
\frac{1}{2}\left(\Delta_R^2 + \Delta_A^2 \right) =
   {\bf P }\left(\frac{1}{p^2}\right)^2
\quad\Leftrightarrow\quad
\Delta_R\Delta_A
\quad\mbox{for}\quad -p^2 \ge k_c^2 > 0 .
 \label{64}
\eeq
With this relation all relevant terms in Eq.~(\ref{63}) proportional to
the fermion distribution $ n(p_0,X)$ are seen to
cancel leaving behind the one loop corrected propagator \cite{our3}
\beq
 \left. \delta S_{12}(p) \right|_{p^2 \le -k_c^2}
   \stackrel{\wedge}{=}
   -\frac{1}{(p^2)^2} \not p \Sigma_{12} \not p.
 \label{65}
\eeq
It has the same form as in the equilibrium case with the only
difference that the quark selfenergy should be calculated with
off-equilibrium distribution function.

\subsection{Quark propagator -  soft  momentum scale}
\bigskip

On the  soft momentum scale, contrary to the discussion in the
last subsection, we can not restrict the calculation
 of the quark propagator
to the one-loop level. Here  as in the equilibrium medium,
  one needs to carry out resummations  to
derive the effective propagator.
In the nonequilibrium case, however,  the
 analogous resummation program is complicated
by the appearance of the extra terms as we have already seen.

\begin{figure}
\centering
\epsfig{file=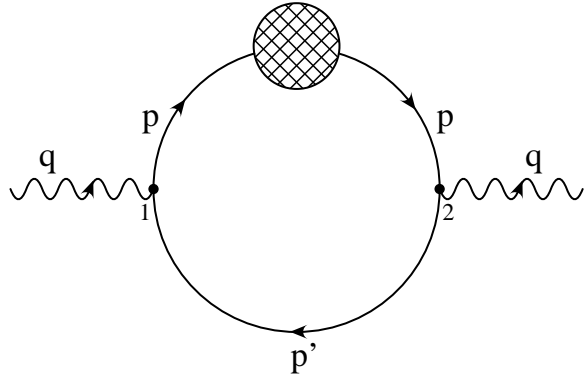,width=8cm}
\caption{\label{fig:res} The photon polarization tensor $\Pi_{12}$
 for soft momentum $p\sim gT$.
 The blob indicates  the HTL-resummed quark propagator}
\end{figure}

The resummed propagator for the off-equilibrium situation
was discussed in \cite{Altherr1,Altherr2} for the scalar case with
the objective of providing a self consistent cut-off for the pinch
contributions.
Generalizing to the
fermionic case resumming successive (one-loop) self-energy insertions
\beq
 iS_{12}^\star (p) = iS_{12}^0 + iS_{1a}^0(-i\Sigma_{ab}) iS^0_{b2} +
  iS^0_{1a}(-i\Sigma_{ab}) iS^0_{bd} (-i\Sigma_{de}) iS^0_{e2} + \cdots
 \label{66}
\eeq
may be rearranged into the effective propagator \cite{our3}, \cite{our4}
\begin{eqnarray}\label{67}
 iS_{12}^\star (p)    &=& -  n_q(p_0,X)
 (\frac{i}{\not p -\Sigma + i\varepsilon p_0} +
        \frac{-i}{\not p -\Sigma^\star - i\varepsilon p_0})    \nonumber \\
 &+&  \frac{i}{\not p - \Sigma + i\varepsilon p_0 }
  \left[ (1- n_q(p_0,X)) (-i\Sigma_{12}) +  n_q(p_0,X) (-i\Sigma_{21}) \right]
  \frac{-i}  {\not p - \Sigma^\star -i\varepsilon p_0},
\end{eqnarray}
making use of the identities
\beq
\Sigma_{11} = - \Sigma_{22}^\star,  \qquad
Im \Sigma = - \frac{i}{2} (\Sigma_{12} - \Sigma_{21}), \qquad
Re \Sigma = Re \Sigma_{11},
 \label{68}
\eeq
where $\Sigma^\star$ denotes complex conjugation and $n_{q(g)}(p_0,X)$
denotes the quark     (gluon) nonequilibrium distribution functions.
The  effective propagator is  well defined
 as it is regularized by the complex retarded quark selfenergy $\Sigma$.

The expression in square brackets of     (\ref{68})
\beq
 (1- n_q(p_0,X)) \Sigma_{12} +  n_q(p_0,X) \Sigma_{21} \equiv
  (1-2 n_q(p_0,X)) \Sigma^- + \Sigma^+,
 \label{69}
\eeq
contains both leading (in the strong coupling $g$)
\begin{eqnarray}\label{410}
 \Sigma^- &=& \frac{1}{2}(\Sigma_{12} - \Sigma_{21}) \sim
 \frac{g^2}{2\pi^2} C_F \int_0^\infty E dE (n_q(E,X) +  n_g(E,X))
  = m_q^2(\lambda_q,\lambda_g), \\
 & & m_q^2(\lambda_q,\lambda_g) = \frac{g^2 T^2}{12} C_F \left( \lambda_g + \frac{\lambda_q}{2}
 \right) = O(g^2 T^2),
\end{eqnarray}
and subleading
\begin{eqnarray}\label{411}
 \Sigma^+ = \frac{1}{2}(\Sigma_{12} + \Sigma_{21})
 &\sim &   p_0 g^2 \int_0^\infty dk k (\frac{\partial}{\partial k}
         n_q(k,X))(1+2n_g(k,X))  = O(g^3 T^2)
\end{eqnarray}
contributions. In the spirit of the HTL-resummation scheme keeping
only leading contributions,  the resummed propagator
can be rearranged into
\beq
 \left. iS_{12}^\star\right|_{HTL} = -\varepsilon(p_0) Re\left(
  \frac{i}{\not p - \Sigma_{HTL}(p) + i\varepsilon} \right),
 \label{412}
\eeq
where pinch-like terms have disappeared.~\cite{our3,our4}

The only  difference of the result in
 (\ref{412}) and  the equilibrium resummed propagator resides  in the
one-loop quark selfenergy which has to be calculated with  modified
distribution functions (\ref{25}).

\subsection{Hard photon rate}
\bigskip

The last step  to derive  the
finite hard photon  rate in off-equilibrium QGP is to calculate
the photon selfenergy in (\ref{62}) with the propagators
derived in (\ref{65}) and in (\ref{412}).
The calculations are
 well described
in the literature for an equilibrium medium \cite{rb}.
Here we quote only the results obtained in \cite{our3,our4}.

The partial rate from the soft part
\beq
 \left. E_\gamma \frac{dR}{d^3 q} \right|_{soft} =
 e_q^2 \frac{3\alpha}{4\pi^3}
  \lambda_q m_q^2(\lambda_q,\lambda_g)  e^{-E_\gamma/T}
  \ln \left[ \frac{k_c^2}{2m_q^2(\lambda_q,\lambda_g)} \right]
 \label{413}
\eeq
is found to be proportional to $m_q^2$,
which also cuts the logarithm
from below as expected. The parameter $k_c\sim \sqrt{g}T$ cutting the
logarithm from above
results from restricting the resummed calculation to soft
exchanged momenta $-p^2 \le k_c^2$.

The hard part calculated with the propagator (\ref{65}) contains
both the singular - cut-off dependent and regular
contributions. The singular part of the rate in the off-equilibrium situation
is obtained  as   \cite{our3}
\beq
  \left. E_\gamma \frac{dR^{sing}}{d^3 q} \right|_{hard} =
    e_q^2  \frac{3\alpha}{4\pi^3}
 \lambda_q m_q^2(\lambda_q,\lambda_g) e^{-E_\gamma /T}  \ln
    \left(\frac{4E_\gamma T}{k_c^2}\right) ,
 \label{414}
\eeq
the regular one as
\begin{eqnarray}
 \left. E_\gamma \frac{dR^{const}}{d^3q} \right|_{hard} &=&
    e_q^2  \frac{2 \alpha \alpha_s}{\pi^4} e^{-E_\gamma /T} T^2 \lambda_q
  ~C(E_\gamma, T, \lambda_q, \lambda_g),
 \label{415}
\end{eqnarray}
with
\begin{eqnarray}
  C(E_\gamma, T, \lambda_q, \lambda_g) &=&  \lambda_q
  \left[ -1 + (1-\frac{\pi^2}{6}) \gamma +
  (1-\frac{\pi^2}{12}) \ln\frac{E_\gamma}{T}  + \zeta_- \right]
  \nonumber\\
  & &\quad + \lambda_q\lambda_g  \left[ \frac{1}{2} - \frac{\pi^2}{8} +
   (\frac{\pi^2}{4} -2) (\gamma+\ln\frac{E_\gamma}{T}) + \frac{3}{2} \zeta'(2)
  + \frac{\pi^2}{12}\ln 2 + (\zeta_+ - \zeta_-) \right]
  \nonumber \\ \label{416}
  & & \quad + \lambda_g \left[ \frac{1}{2} + (1-\frac{\pi^2}{3})\gamma +
   (1-\frac{\pi^2}{6} )\ln\frac{E_\gamma}{T} - \zeta_+ \right].
\end{eqnarray}
The appearing symbols are Euler's constant $\gamma$, the derivative of the
Riemann $\zeta$-function evaluated at $\zeta'(2)$ and the sums
\begin{equation}
 \zeta_+ = \sum_{n=2}^\infty \frac{1}{n^2} \ln(n-1) \simeq 0.67, \qquad
 \zeta_- = \sum_{n=2}^\infty \frac{(-)^{n}}{n^2} \ln (n-1) \simeq -0.04.
 \label{417}
\end{equation}
The result Eq.(\ref{415}) is successfully checked for consistency
by noting that the constants present in the equilibrium result
\cite{rb} are reproduced for $\lambda_q=\lambda_g=1$.

Calculating the complete emission rate for real photons as the
sum of the two partial rates Eqs.(\ref{413},\ref{414})   and
(\ref{415}),
\begin{equation}\label{eq:reshard}
 E_\gamma \frac{dR}{d^3 q}  =
 \left. E_\gamma \frac{dR^{}}{d^3 q}
  \right|_{soft} +
 \left. E_\gamma \frac{dR^{sing}}{d^3 q} \right|_{hard} +
 \left. E_\gamma \frac{dR^{const}}{d^3 q} \right|_{hard}.
 \label{418}
\end{equation}

We see that the dependence on the arbitrarily chosen $k_c$ cancels and
the final result  for the hard photon rate can be written as
\begin{equation}\label{eq:resc}
  E_\gamma \frac{dR}{d^3q} = e_q^2 \frac{\alpha\alpha_s}{2\pi^2} \lambda_q T^2
   e^{-E_\gamma/T} \left[ \frac{2}{3}(\lambda_g + \frac{\lambda_q}{2})
  \ln \left(\frac{2E_\gamma T}{m_q^2(\lambda_q,\lambda_g)} \right)
 + \frac{4}{\pi^2} C(E_\gamma ,
  T, \lambda_q, \lambda_g) \right] .
 \label{419}
\end{equation}

The above result shows  that as in the equilibrium medium
the generalized thermal mass is
established as self consistent cut-off for the logarithmic singularity.
The dynamical screening of the mass singularity seen in (\ref{419})-here
 given for the distributions (\ref{26a})
actually does not depend on the explicit form of the
nonequilibrium quark and gluon distribution functions.
 Changing the parameterization
of the  distribution functions enters only through the redefinition
of the mass parameter in (\ref{412})   and the constant $C$ in (\ref{416}),
keeping at the same time the functional form of the rate unchanged.

\subsection{Photon production in an expanding quark gluon plasma}
\bigskip

In the following we discuss the
importance of nonequilibrium
effects on hard photon production. Having in  mind the possible
phenomenological implications of the results we
 include the space-time evolution of the
off-equilibrium quark gluon plasma. We start with the assumption
that the partonic system produced in     heavy ion
collisions achieves local thermal equilibrium after  the time
$\tau_i$. Deviations from chemical equilibrium of quarks
and gluons are measured initially by the fugacities $\lambda_q^i$ and $\lambda_g^i$.
Beyond this point the system expands according to
the Bjorken model \cite{bjor} and the chemical equilibration, that is the time dependence
of fugacities is governed by rate equations.

If one assumes the off-equilibrium plasma to undergo only
longitudinal expansion then according to the hydrodynamical equation
the time $\tau$-dependance of temperature is governed by the well known
Bjorken's relation \cite{bjor}, (valid for a free gas)
\beq
\epsilon \tau^{4/3}=const ,
 \label{420}
\eeq
where for the energy density  of a nonequilibrium plasma we take
\beq
\epsilon =({{8\pi^2}\over {15}}\lambda_g+{{14\pi^2 n_f}\over {40}}
\lambda_q)T^4,
 \label{421}
\eeq
where $n_f\sim 2.5$ is  the number of dynamical quark  flavours.

For                the approach to chemical equilibrium
we                                include only the dominant
reaction mechanism for parton production. These are just
four processes
\beq
gg <->ggg~~,~~gg<->q\bar q.
 \label{422}
\eeq
With the above processes
the rate equation can be written as
 the following master equations in
terms of the weighted reaction rates $R_{2,3}$ \cite{biro}:
\beq
{1\over {\lambda_g}}{{d\lambda_g}\over {d\tau}}
+
{3\over {T}}{{dT}\over {d\tau}} +
{1\over {\tau}} =R_3(1-\lambda_g)-2R_2
(1-{{\lambda_q^2}\over {\lambda_g^2}}) ,
 \label{423}
\eeq
$$
{1\over {\lambda_q}}{{d\lambda_q}\over {d\tau}}
+
{3\over {T}}{{dT}\over {d\tau}} +
{1\over {\tau}} ={{32}\over {9n_f}}R_2
({{\lambda_g}\over {\lambda_q}}
-{{\lambda_q}\over {\lambda_g}})  ,
$$
describing together with (\ref{420}) the time evolution of the parameters $T(\tau )$, $\lambda_q(\tau  )$
and $\lambda_g(\tau  )$.

\begin{figure}
\centering
\epsfig{file=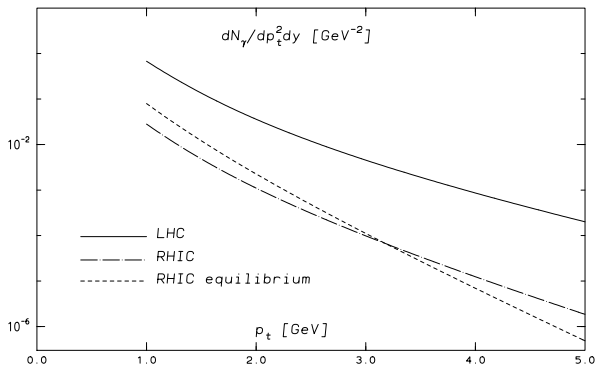,width=8cm}
\caption{\label{fig:kn145} Distribution of thermal photons for $y=0$
 from off-equilibrium QGP for the conditions at RHIC and LHC
 energies. The equilibrium result for RHIC is shown by the dotted line }
\end{figure}

In Fig.~9 we calculate the space-time integrated nonequilibrium
photon rate (\ref{419}).
The initial conditions have been fixed for RHIC:
$\tau_i\sim 0.25{\rm fm},~T_i\sim 0.67GeV,~\lambda_q^i\sim 0.064,~\lambda_g^i\sim0.34$ and LHC:
$\tau_i\sim 0.25{\rm fm},~T_i\sim 1.02GeV,~\lambda_q^i\sim 0.082,~\lambda_g^i\sim0.43$
as suggested by the Self-Screened Parton Cascade model \cite{ss}.

To emphasize the nonequilibrium effects we also show in Fig.~9
the photon rate for  the RHIC experiment calculated with equilibrium distributions
of  quarks and gluons. The initial temperature $T_i\sim 0.43$GeV is fixed
requiring the initial energy density $\epsilon_i\sim$61.4GeV/fm$^3$.
 The results in Fig.~9 indicate the  differences between
 equilibrium and off-equilibrium photon yields. As expected the
 higher initial temperature for the chemical off-equilibrium
 scenario increases the yield of large $p_t>3$GeV photons.

Turning finally to comparisons of our results with previous studies
  we note first that the rate
(\ref{419}) differs in details from the results of both
 Refs.\cite{s1} and
\cite{s2,s3}; but
indeed justifies the cut-off prescription employed therein. The
discrepancy with respect to
\cite{s2} can be attributed to the fact
that  distributions for the incoming particles
have been treated in Boltzmann approximation. We find, however, that it is
necessary to keep track of full quantum statistics in order for the screening
to be complete.

 The photon yield calculated in Fig.~9 also differs from the
results given in \cite{d1}. This is  related to  the new structure of the
off-equilibrium photon rate derived in our approach. However, although
our estimate  shows a different
functional dependence of the photon rate on the  fugacity parameters,
nevertheless the numerical differences with previous estimates are small
 being of the order of 20$\% .$
\section{Summary}
\bigskip

We have calculated the thermal production rate for soft
dielectrons produced in off-equilibrium  pionic medium including  relevant
 reactions from $\pi-\rho$ interactions.
 The calculations include the  $O(g_{\rho}^2)$ corrections
 arising from the two-loop contributions to the virtual photon
selfenergy.
We have discussed the importance of the pinch term
which arises from the structure of the nonequilibrium effective
 particle propagator.
The appearance of the pinch  term modifies substantially the
 properties   of the photon  rate
leading e.g. to  the enhancement of the soft dilepton production 
 in an undersaturated medium.

The second important result of our analysis is that we show explicitly
the absence of
any additional contributions from possible pinch (singular) terms to
the off-equilibrium photon production rate in a quark gluon plasma.
As far as the soft momentum scale is concerned
these terms are shown to be subleading with respect to the
dominant hard thermal loop contributions. On the hard scale on the other hand
the absence of pinch singular contributions is due to
the restricted kinematics of real photon production.
This result
does not
apply in general, e.g., for virtual photon production
the relevant region  of phase space is sufficiently enlarged for
these terms to become substantial. This has been  shown to hold in our
estimate of dilepton production from a mesonic medium \cite{our1}
 and it  is also valid
in the case of  the quark gluon plasma \cite{LeBellac2}.

\section*{Acknowledgements}

Work  supported in part by the Deutsche Forschungsgemeinschaft (DFG).
One of us K.~R. acknowledges
 partial support by Zentrum f\"ur interdisziplin\"are Forschung (ZIF) and
 stimulating discussions with F.~Karsch,
H.~Satz and D.K.~Srivastava.
We   acknowledge   discussions with D. Schiff.
%
%

\end{document}